\documentclass[pdflatex,sn-mathphys-num]{sn-jnl}
\usepackage{graphicx}%
\usepackage{multirow}%
\usepackage{amsmath,amssymb,amsfonts}%
\usepackage{amsthm}%
\usepackage{mathrsfs}%
\usepackage[title]{appendix}%
\usepackage{xcolor}%
\usepackage{textcomp}%
\usepackage{manyfoot}%
\usepackage{booktabs}%
\usepackage{algorithm}%
\usepackage{algorithmicx}%
\usepackage{algpseudocode}%
\usepackage{listings}%
\DeclareUnicodeCharacter{394}{$\Delta$}

\DeclareMathOperator*{\argmax}{arg\,max}

\begin{document}

\title[LFaB]{LFaB: Low fidelity as Bias for Active Learning in the chemical configuration space}

\author*[1]{\fnm{Vivin} \sur{Vinod}}\email{vinod@uni-wuppertal.de}
\author[1]{\fnm{Peter} \sur{Zaspel}}
\affil[1]{\orgdiv{School of Mathematics and Natural Sciences}, \orgname{University of Wuppertal}, \orgaddress{\country{Germany}}}

\abstract{
Active learning promises to provide an optimal training sample selection procedure in the construction of machine learning models. It often relies on minimizing the model's variance, which is assumed to decrease the prediction error. Still, it is frequently even less efficient than pure random sampling. Motivated by the bias-variance decomposition, we propose to minimize the model's bias instead of its variance. By doing so, we are able to almost exactly match the best-case error over all possible greedy sample selection procedures for a relevant application. Our bias approximation is based on using cheap to calculate low fidelity data as known from $\Delta$-ML or multifidelity machine learning. We exemplify our approach for a wider class of applications in quantum chemistry including predicting excitation energies and \textit{ab initio} potential energy surfaces. Here, the proposed method reduces training data consumption by up to an order of magnitude compared to standard active learning. 
}

\maketitle

\section*{Main}\label{sec_intro}

Machine learning (ML) surrogate models are routinely applied to replace the evaluation of computationally expensive simulation models, such as in quantum chemistry (QC)
\cite{behler2015constructing, butler_davies_review_2018, baum_2021_AIChemreview, dral2020quantum, schwaller_ML_inchem_review}, drug discovery \cite{schneider2010virtual,SCHAPIN2023100020}, or beyond. Once trained, these models offer rapid \textit{predictions}, while the computational cost is hidden in the time to generate the training data. Since each training sample is computationally expensive, it is natural to minimize the required training data to maximize the model's total cost efficiency.

Active learning (AL) strategies are a family of greedy training sample selection procedures that should support the minimal use of training data. There exist several heuristics in AL for the selection of training samples such as low-confidence sampling \cite{white1992handbook} or highest-change sampling \cite{atlas1989training}. A common low-confidence approach uses \textit{uncertainty sampling}. This procedure sequentially selects and labels input samples from a large pool of unlabeled candidate samples based on an \textit{uncertainty measure}. Usually, the selected uncertainty measure is some approximation to the model's variance, leading to a greedy minimization of the model's variance. Various such variance approximations exist: Gaussian Process Regression (GPR) provides an estimate to the model's variance as part of the predictor \cite{williams2006gaussian}. Ensemble-based variance estimates \cite{rupp2014machine, uteva2018_activelearning, guan2018construction, Vandermause_2020_onthefly_AL, christiansen2024efficient} build an ensemble of models on subsets of the training data and estimate the variance from the ensemble's output. Also model variance with respect to changing model architectures, while keeping identical training data, can be considered \cite{Huang_Hou_Dral_2025_deltaAL}. 

While the use of AL techniques for ML models in e.g.~excitation dynamics calculations \cite{Bispo_2025_MELTS, Garain_Jr_Bispo_Barbatti_2025_uncertainty} of QC shows very favorable results in stabilizing the dynamics, its use in settings, where cost-efficient training is of highest importance, is debatable. To be more specific, we consider a setting in which a larger set of non-labeled (inputs of) training samples is given. Then the task is to adaptively select, label and include more and more training samples based on a sample selection strategy. Despite many methodological improvements, AL approaches, e.g.~in QC, are in this setting often not capable to outperform even a very simplistic pure random sampling, in terms of the error for growing number of samples \cite{holzenkamp2025_uncertainty_GPR,Garain_Jr_Bispo_Barbatti_2025_uncertainty}. Roughly speaking, variance-based sample selection strategies tend to favor samples that are in regions of the sample space that are under-resolved by the concretely selected training subset. This would indeed lower the variance of the model. Still, empirical results show that this lowered variance does not immediately imply a smaller prediction error. 

In this context, it is worth while recalling that the prediction error of a machine learning model for perturbed input data is the sum of a contribution from the perturbation of the data, from the variance and from the \textit{bias}, as known from the bias-variance decomposition \cite{Geman1992NeuralNA,Domingos2000AUB}. Our working assumption is that, when AL methods fail to lower the prediction error while using variance information for sample selection, the prediction error is dominated by the error contribution of the bias and not by the error contribution of the variance. 

To overcome this limitation, we hence propose to use information on the bias of the model, rather than the variance of the model, to guide the process of training sample selection and labeling in AL for the depicted challenging cases. Calculating the actual bias of the model for training sample selection would require to have the true error of the model, hence the labels for all samples of the candidate training sample pool. Instead of doing so, we propose to \textit{approximate} the bias, by using labels that are much cheaper to compute than the costly labels that are to be adaptively selected. That is, the use of a lower fidelity to compute approximate labels. This gives rise to a strategy that uses labeling with two different types of labels, one expensive to calculate but accurate and one cheap to calculate but still a ``reasonable'' approximation of the expensive label. In this work, we refer to this as the Low-Fidelity-as-Bias (LFaB) method. 

We exemplify this strategy for the application field of QC, particularly for the construction of machine learning models for excitation energies and \textit{ab initio} potential energy surfaces (PES). In this field, a combined use of cheaper and more expensive labels has been proposed in context of $\Delta$-ML \cite{Ramakrishnan2015} and multifidelity-type machine learning models \cite{zasp19a, vinod2024_benchmarking_dataefficiency_deltaml}. While conceptually, the utilized data is similar to these methods, the here proposed method uses the cheap data to guide the sample selection process, instead of acting immediately on model output. 

In the following results section, we will not only show that this bias-based AL clearly outperforms a random sample selection strategy and several typical variance-based sample selection strategies, it often even almost perfectly matches the best-case greedy-optimal approach for selecting samples \cite{schaback2000adaptive, wirtzhaasdonk2013_greedy_selection,ling2009improved}, which would use a fully (expensively) labeled candidate training set to select the next best sample to pick. 

Benchmark experiments indicate that the use of our new strategy reduces the number of AL iterations required to reach a specific empirical error by a factor of 5 compared to the use of variance-based AL in a certain case of predicting excitation energies and \textit{ab initio} PES. In cases such as predicting atomization energies, it is seen that the our approach reduces empirical error by a factor of about 2. We also show that the bias information leads to quite optimal results in a so-called calibration analysis commonly made for the analysis of AL schemes \cite{kuleshov_accurate_2018, levi_2022_UQ_calibration}. Overall, this work provides strong reduction in the overhead of the use of ML exemplified for the field of QC. Since the novel scheme is also simple to implement, it is a ready-to-use tool for researches working in the ML-QC field. 

\section*{Results}\label{sec_results}
\begin{figure}[H]
    \centering
    \includegraphics[width=\linewidth]{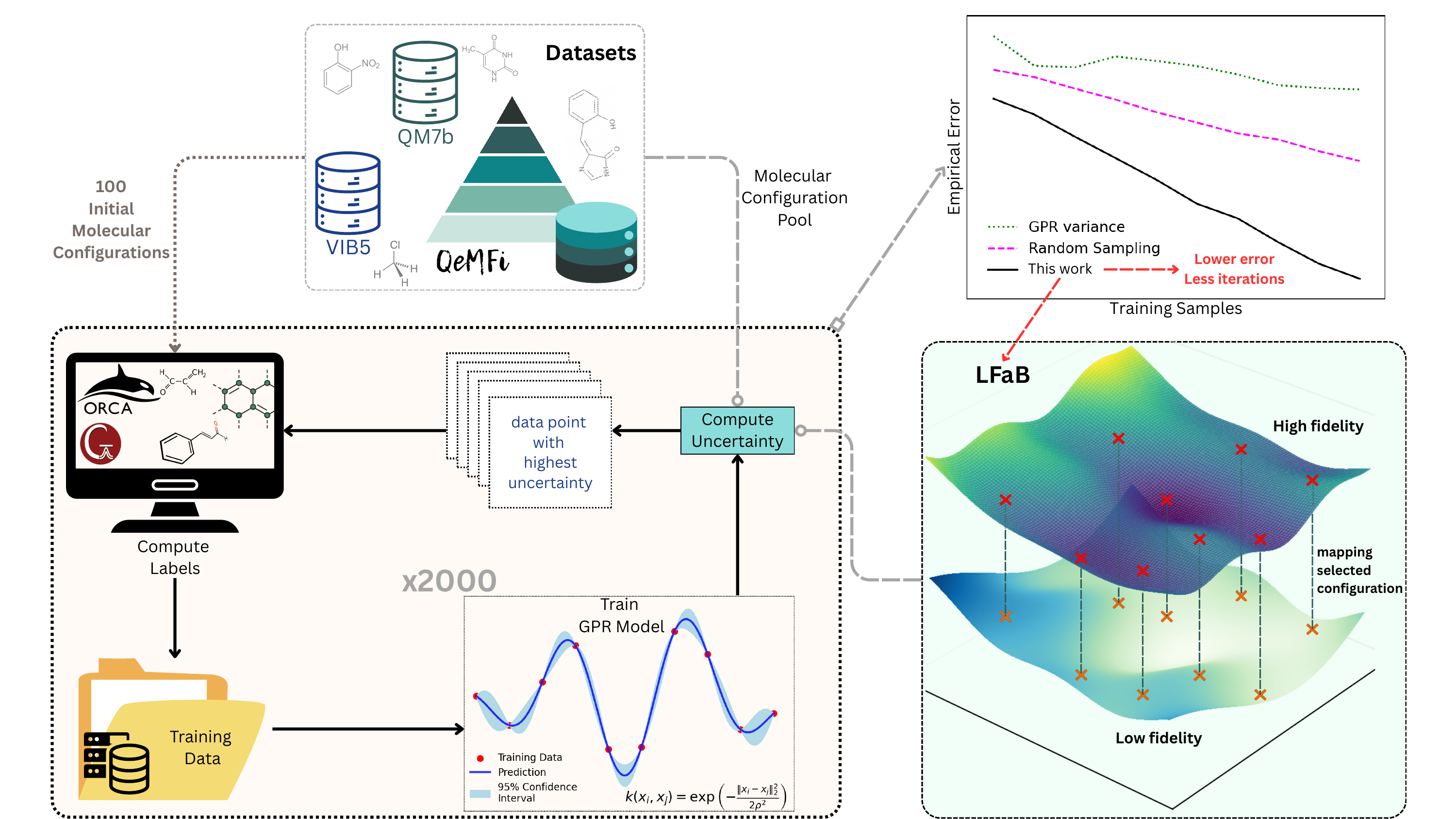}
    \caption{A pictorial representation of the workflow employed to carry out numerical benchmarks of the low fidelity as bias (LFaB) method. The numerical benchmarks are performed for three datasets, QM7b \cite{montavon2013machine}, VIB5 \cite{zhang_vib5_2022}, and QeMFi \cite{vinod2024QeMFi_paper} for atomization energies, \textit{ab initio} potential energy surfaces, and excitation energies respectively. An initial set of molecular configurations is chosen to train the GPR model after making computations of the QC properties (or labels). The uncertainty of the trained model, either the variance or the bias, is estimated for the unlabeled molecular configurations in the training data pool. As seen at the lower right corner, LFaB uses the lower fidelity as a measure of the bias of the model trained at the higher fidelity. In this work, the LFaB scheme is benchmarked against alternative sampling techniques in active learning based on GPR model variance, variance of model ensembles, and against a greedy-optimal selection.       
    }
    \label{fig_LoUQAL_concept}
\end{figure}

The workflow of AL in QC is depicted in Fig.~\ref{fig_LoUQAL_concept}. The starting point is a collection, or data pool, of molecular configurations. From the data pool an initial training set is generated by randomly selecting $n_{\rm init}$ configurations and computing their labels. This initial training dataset is used to train the ML model, $g_{ML}$. The uncertainty of this model is computed either as variance or the bias over the data pool. Next, the molecular configuration from the pool with the highest uncertainty is selected and the label computed. This pair is then appended to the training dataset followed by retraining the model $g_{ML}$. This procedure is repeated iteratively until a satisfactory threshold of either uncertainty or training data set size is reached. In this work, 2000 iterations of active learning are carried out with one sample being added each iteration. 

Variance based uncertainty of the trained model include the use of GPR variance (Eq.~\eqref{eq_gpr_variance}), or the variance of predictions of an ensemble of ML models. 
One bias based uncertainty of the ML model is the greedy-optimal sampling which uses the difference between the prediction from $g_{ML}$ to the actual value of the label. As mentioned above this would require the prohibitive computation of the labels for the entire data pool. In contrast, random sampling of training samples from the data pool, which is the sampling approach used in ML without active learning, randomly picks molecular configurations from the data pool for which the labels are computed and then added to the training data. 
The reader is directed to \nameref{sec_uncertainty_measure} for further details on the different methods of sampling training data from the data pool.

On the other hand, as depicted in the lower right corner of Fig.~\ref{fig_LoUQAL_concept}, the LFaB method utilizes reference labels computed at a lower fidelity and a ML model trained at the lower fidelity to compute the bias. In this approach, first a low fidelity computation of the labels is made for the entire data pool. Next, we select $n_{\rm init}$ samples to form the initial training dataset. For the molecular configurations chosen, the labels at the high fidelity (that is, the fidelity that one is actually interested in) are also computed. The approximation of the bias is made using the difference in prediction of the low fidelity model to the low fidelity labels. The molecular configuration with the highest bias is selected and the low fidelity and high fidelity labels are computed. The pair of molecular configuration and label are added to the respective training dataset. The details of the algorithm are given in \nameref{lfab_method}.

In this work, three distinct QC properties - atomization energies of the QM7b dataset \cite{montavon2013machine}, \textit{ab initio} potential energy surfaces of the VIB5 dataset \cite{zhang_vib5_2022}, and excitation energies of the QeMFi \cite{vinod2024QeMFi_paper} dataset - are chosen in order to thoroughly benchmark the proposed LFaB method. We utilize GPR as the ML models in this work (see \nameref{sec_GPR_methods}).
For each of these properties, the available data is first partitioned into a 90/10 split of \textit{training pool} and holdout \textit{test set} data. The latter is used to compute the empirical error of the trained GPR model, reported as mean absolute error (MAE) which is given by Eq.~\eqref{eq_MAE}. In this work, each experiment uses $n_{\rm init}=100$. Hyper-parameters of the model $g_{ML}$ are optimized on the initial training dataset using marginal log-likelihood. 
In the numerical experiments below, the LFaB method is compared against AL that uses GPR variance and model ensemble variance to select samples each iteration. In addition, we compare the use of randomly sampled data and the best-case greedy-optimal approach.

\subsection*{LFaB benchmark for atomization energies of QM7b}
QC methods offer a clear hierarchy in terms of computational cost and achieved accuracy of calculation. The Hartree Fock (HF) method is considered cheaper and less accurate than Møller-Plesset perturbation (MP2) which is cheaper and less accurate than the coupled cluster singles doubles triplet (CCSD(T)) method \cite{jens17a}. In this experiment, we benchmark the use of LFaB for the prediction of atomization energies in the QM7b dataset \cite{montavon2013machine} where the fidelity of data is ordered by the choice of the QC method used. 
The QM7b dataset \cite{montavon2013machine} consists of atomization energies of a total of 7,211 molecules computed with 3 different QC methods CCSD(T), MP2, HF. For each, three basis sets are used, STO3G, 631G, and ccpVDZ (see section \ref{sec_qm7_data}). Here, we fix the choice of the basis set and choose CCSD(T) to be the high fidelity label. Subsequently, MP2 and HF are then selected as the lower fidelity. The molecules are represented using SLATM descriptor \cite{Huang2020slatm}. 

\begin{figure}[H]
    \centering
    \includegraphics[width=\linewidth]{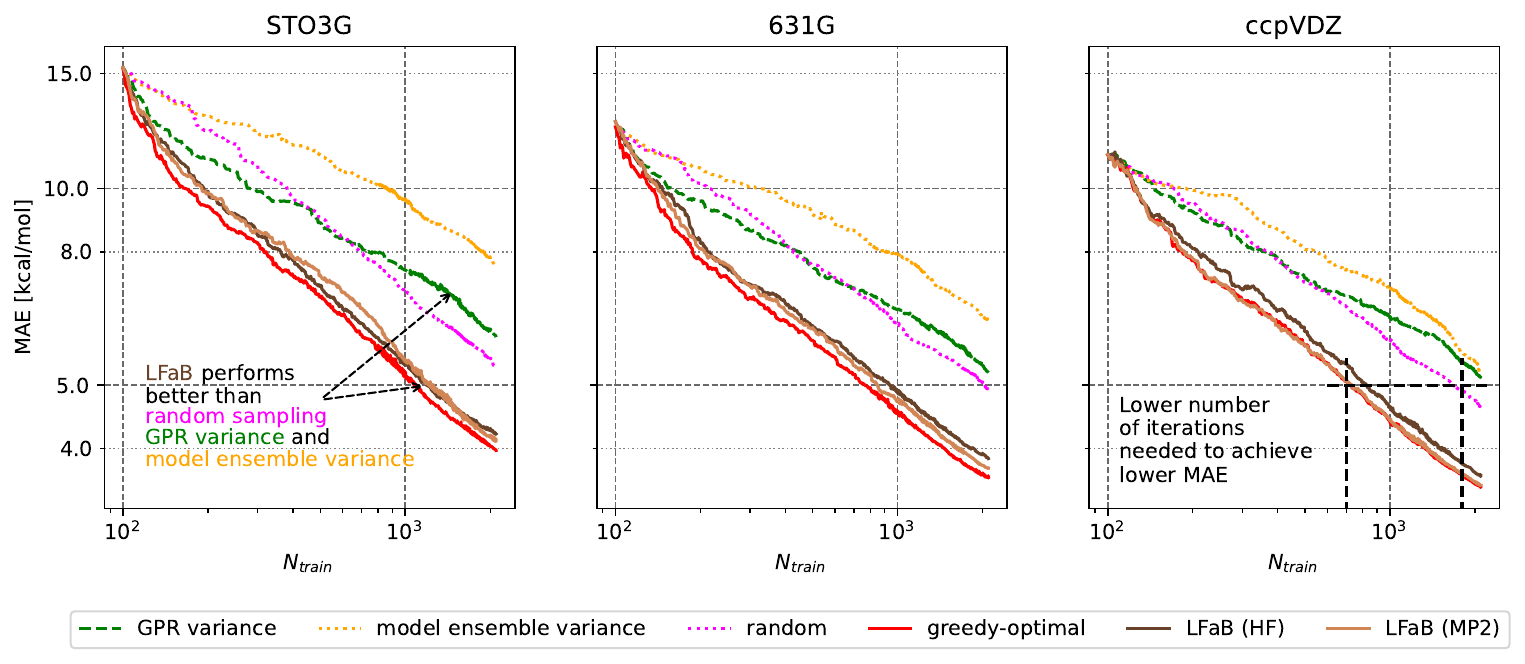}
    \caption{Learning curves for the prediction of atomization energies of the QM7b dataset with different sampling techniques for active learning. For the LFaB scheme, the low fidelities are expressed by the choice of the QC method with each pane showing a different basis set choice. Thus, the high fidelity is CCSD(T) with the low fidelity being either HF or MP2. Reference dashed lines in the right-hand side plot establish that the use of the LFaB sampling scheme reduces the number of active learning iterations required to achieve a certain error in comparison to the random selection of training samples. The LFaB method is seen to be superior to other variance based sampling techniques.}
    \label{fig_QM7b}
\end{figure}

Resulting learning curves are shown in Fig.~\ref{fig_QM7b}. In all three panes, the horizontal axis reports number of training samples, each chosen during a specific run of the AL iteration, while the vertical axis reports MAE on the holdout test set. Both axes are scaled logarithmically. Consider the case for basis set ccpVDZ. The learning curve corresponding to model ensemble variance performs the worst in this case with a higher error throughout all training set sizes. Next comes the one for GPR variance with an error above the 5 kcal/mol mark for 2,000 adaptively selected samples. Both these sampling schemes perform worse than randomly sampling training data across the configuration space. 
This indicates that these sampling techniques do not result in a substantial reduction of empirical error with the addition of the sample with the highest uncertainty. This was also previously reported in ref.~\cite{holzenkamp2025_uncertainty_GPR} for the development of ML interatomic potentials. 
On the other hand, the greedy-optimal sampling results in a learning curve that is steeper and results in a much lower error. This is anticipated since, here, the full knowledge of the atomization energy landscape at the target fidelity is known. The learning curve for the LFaB method is reported with the low fidelity as the QC method in parenthesis in Fig.~\ref{fig_QM7b}. 
With either HF or MP2, The LFaB sampling scheme performs nearly identical to the greedy-optimal method with a similar slope resulting in an MAE of about 3 kcal/mol for 2,000 adaptively selected samples. Identical observations can be made for the other two basis sets (left and center pane of Fig.~\ref{fig_QM7b}) where the LFaB method outperforms random sampling and the existing sampling schemes. 

One inference from these learning curves can be readily made about LFaB vis-\'a-vis other sampling schemes.  
For a fixed number of AL iterations, the LFaB method outperforms all other sampling techniques in terms of empirical error. Alternatively, if one had a threshold of error that was desired, say 5 kcal/mol, then the LFaB method achieves this error for a fewer number of AL iterations ($\sim700$) than the random sampling approach ($\sim1,800$) while the use of GPR variance and model ensemble variance to select samples will likely reach this error with more than 2,000 iterations. In this particular case, the use of LFaB results in use of \textit{60\% less training data} in comparison the the use of random selection to achieve the same MAE. 
The negative slope of the learning curves for LFaB indicates that the addition of further samples would decrease the MAE of the trained GPR model. 

\subsection*{LFaB benchmark for \textit{ab initio} energies of VIB5}
The VIB5 database provides the ground state \textit{ab initio} PES for different molecules. For each molecule, several configurations are given with the corresponding value of the PES being computed with varying methods, the most expensive being CCSD(T) with the ccpVQZ basis set \cite{zhang_vib5_2022}. Two such molecules are $\rm CH_3Cl$ and $\rm CH_3F$.
Lower fidelities are defined by both cheaper QC methods and  smaller basis set sizes. Thus, the use of this dataset for benchmarking the LFaB method would involve the definition of fidelities in a composite manner - as a mix of QC method and basis set size. 
Since there are tens of thousands of configurations for each molecule making the dataset rather large, in this experiment, for each of $\rm CH_3Cl$ and $\rm CH_3F$, 15,000 samples were randomly chosen and a 90/10 split into training pool and test set was carried out. Since the experiments were carried out for the same molecule and no cross molecule tests were carried out, the unsorted CM matrix representation \cite{RupCM} was used without needing to account for permutational invariance since the atoms are numbered identically across the entire dataset of each molecule in VIB5 (see \ref{sec_representations}). 
The higher fidelity is the gold standard CCSD(T) method used with the ccpVQZ basis set.
The lower fidelities used in this case are MP2-ccpVTZ, HF-ccpVQZ, and HF-ccpVTZ. In the forthcoming results, shorthand notations such as HF-TZ are used. 

\begin{figure}[H]
    \centering
    \includegraphics[width=0.8\linewidth]{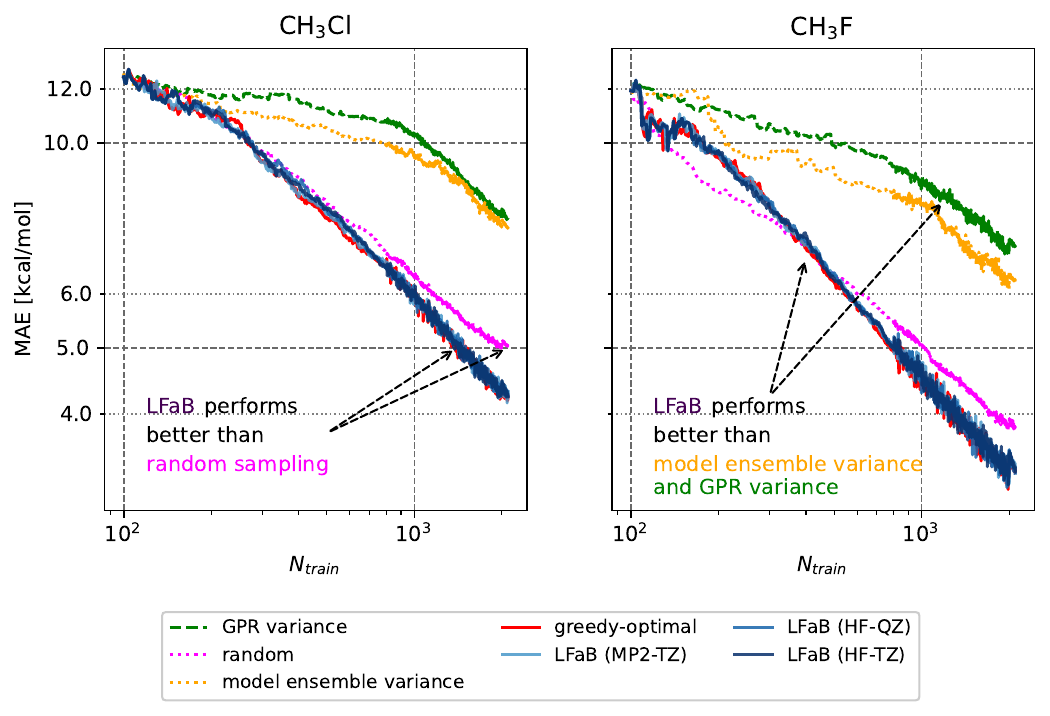}
    \caption{Learning curves showing prediction error (measured as MAE) versus number of training samples for the prediction of \textit{ab initio} PES for $\rm CH_3Cl$ and $\rm CH_3F$ from the VIB5 database.
    Different active learning sampling schemes are used for each learning curve shown. The LFaB method outperforms the use of random sampling of training data in terms of prediction error. Furthermore, it results in a prediction error that is near identical to the greedy-optimal sampling approach outperforming both GPR variance and model ensemble variance.}
    \label{fig_VIB5}
\end{figure}

Learning curves for the AL sampling strategies in predicting \textit{ab initio} PES for two molecules of the VIB5 database are shown in Fig.~\ref{fig_VIB5}. Consider the case for $\rm CH_3Cl$ on the left-hand side of the figure. 
The model that uses GPR variance to sample new points for the training dataset performs poorest showing a gradual slope of the learning curve. A similar observation is made for the model ensemble variance sampling scheme with both resulting GPR models showing an MAE around 8 kcal/mol for around 2,000 training samples. The random sampling approach results in a steeper slope than using either of these sampling schemes indicating that both these schemes lack performance. On the other hand, the LFaB scheme performs better than the random sampling approach and results in an MAE as good as the greedy-optimal approach. The model built with LFaB sampling with either of the three lower fidelities perform similar to each other, still better than the random sampling approach, with no apparent difference in the overall empirical error arising from the final models. 
The negative slope of the learning curves for LFaB indicate that addition of training samples could further reduce the error of the trained GPR model.
In comparison to the model built with GPR-variance used for sampling, the model using LFaB results in a lower MAE, almost a \textit{factor of 2} for $\rm CH_3Cl$. In terms of training samples, it can be seen that for a MAE of around 8 kcal/mol, the model built with GPR variance for sampling requires 2,000 iterations while the model using LFaB sampling achieves this error with around 400 iterations. Since the training dataset is appended one sample at a time, this directly correlates to the number of QC calculations. Therefore, it can be seen that the LFaB approach reduces the training data cost (and the AL iterations) by \textit{a factor of 5} in this case. 

Another test of defining fidelities as a combination of QC method and basis set choice is performed in the supplementary information section \ref{sec_composite_QM7b} for the QM7b dataset as well. The resulting learning curves are shown in Fig.~\ref{fig_QM7b_fullrange}. In this case as well, the LFaB method outperforms the random sampling of configuration space while being close to the accuracy offered by the greedy-optimal sampling scheme. With the high fidelity being CCSD(T)-ccpVDZ, one notices that the choice of the basis set makes more of a difference than the choice of the QC method while employing the LFaB sampling method for AL.

\subsection*{LFaB benchmark for excitation energies of QeMFi}
This section benchmarks the prediction of excitation energies with fidelities being defined as the choice of basis set while fixing the QC method to be DFT. As briefly mentioned above, in Fig.~\ref{fig_QM7b_fullrange} it was observed that the choice of the basis set made a stronger difference than the choice of the QC method while predicting the atomization energies. Therefore, it is relevant to check whether this effect is visible in the case of a more complex QC property, the excitation energy.  The excitation energy of molecules is in general known to be a more complex property for ML models to learn \cite{Westermayr2020review, dral_molecular_2021} and therefore is an additional challenge for the herein proposed LFaB method.
The QeMFi dataset contains 9 molecules with 15,000 configurations each. For each configuration, excitation energies are computed with the DFT formalism using varying basis set sizes \cite{vinod2024QeMFi_paper, vinod_2024_QeMFi_zenodo_datatset}. The high fidelity is def2-TZVP (hereon denoted as TZVP) with the lower fidelities being one of the smaller basis sets: def2-SVP (hereon denoted as SVP), 631G, and 321G. The benchmarks are carried out for each individual molecule and therefore the unsorted CM \cite{RupCM} are used not needing to account for permutational invariance. 

\begin{figure}[H]
    \centering
    \includegraphics[width=\linewidth]{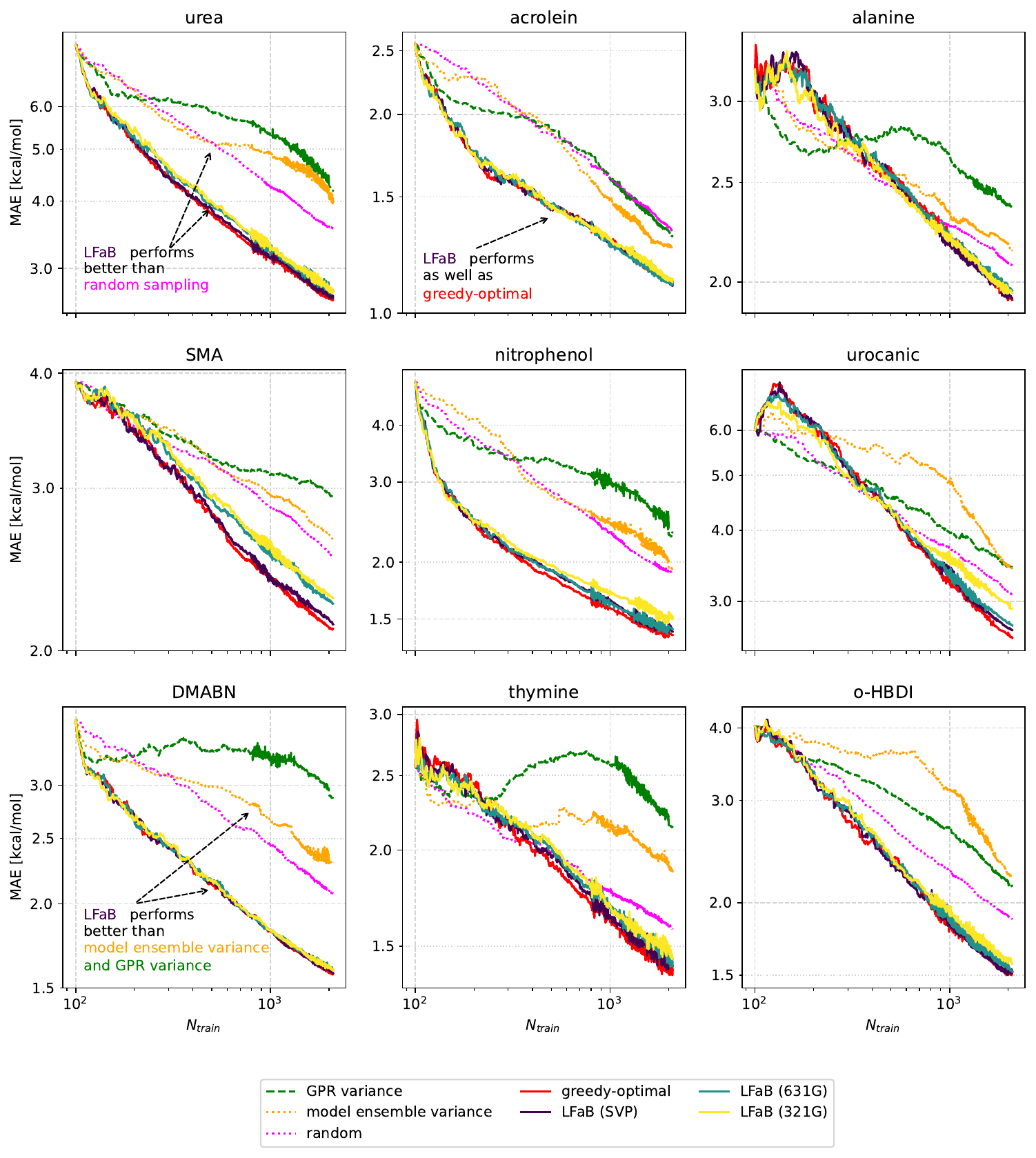}
    \caption{Learning curves for the prediction of excitation energies of diverse molecules from the QeMFi dataset. Each data point is added after a cycle of active learning with the corresponding sampling scheme being used. The terms in the parenthesis for the LFaB measure indicate the lower fidelity used, in this case corresponding to the basis set chosen to make the quantum chemistry calculation. 
    The random sampling approach is also shown for contrast, being the common approach that is followed in general machine learning in quantum chemistry workflows. In every case, the LFaB method performs better than the random sampling approach and other active learning sampling techniques.}
    \label{fig_qemfi_excitation}
\end{figure}

Learning curves for the prediction of excitation energies of each molecule in QeMFi are shown in Fig.~\ref{fig_qemfi_excitation} once again depicting the different AL sampling schemes studied in this work alongside the random sampling approach. 
For each of the nine molecules identical observations can be made. The models built with the GPR variance and model ensemble variance sampling methods perform worse than the random sampling of configuration space. This implies that one is better off not employing AL with these sampling schemes if the sampling is performed with existing variance based methods.
On the other hand, the model that uses LFaB to sample training data results in learning curves with much steeper slopes, almost same as the greedy-optimal sampling method. 
This once again indicates that the novel LFaB method results in a trained ML model that reduces the empirical error of prediction performing better than the random sampling approach and existing methods of sampling that employ variance to select new samples. Consider for instance the case for DMABN in Figure \ref{fig_qemfi_excitation}. Here, the model using GPR variance to sample training data results in a MAE around 3 kcal/mol even after 2000 AL-iterations. For the same number of iterations the model built using the LFaB scheme results in an MAE of about 1.5 kcal/mol. That is, a model that is twice as accurate.
An alternative way to study these curves is to fix a desired MAE from the ML model. Say one wished a ML model with MAE of 2.5 kcal/mol. The model built using GPR variance to sample training data does not reach this error. The model built with model ensemble variance reaches this error after 2,000 iterations. The LFaB method results in a model with this error with as little as 400 iterations indicating a reduction in number of iterations (and thereby training samples) by a \textit{factor of 5}. Thus for the prediction of excitation energies, the LFaB method is a powerful tool in reducing the number of training samples without compromising on model accuracy. In this specific experiment, it is seen that the choice of basis sets as the fidelity hierarchy does not reduce the effectiveness of the LFaB method. The negative slope of the learning curves for LFaB denote that the addition of more training samples will reduce the MAE of the model. In the supplementary information section \ref{SI_batch_addition} it is shown that the addition of training samples in batches using the LFaB method converges to the same MAE as the case of single point addition that is carried out in this work. 

\subsection*{Visualizing Uncertainty} \label{sec_PCA_results}
Having benchmarked the LFaB method for different QC properties, it is useful to also establish its credibility in a visual comparison to the greedy-optimal scheme. The learning curves from the previous sections already provide numerical evidence of the performance of the LFaB scheme. Since the use of ML in QC first requires the conversion of molecular configurations to representations/descriptors, it is relevant to look at this as a proxy of the chemical configuration space. To this end, the molecular descriptors are passed through a principle component analysis (PCA) with the two leading components being used to visualize the proxy space. The working hypothesis is that any good sampling technique for active learning would select those points in the chemical configuration space that are nearly the same as the ones picked by the greedy-optimal sampling scheme.

\begin{figure}[H]
    \centering
    \includegraphics[width=\linewidth]{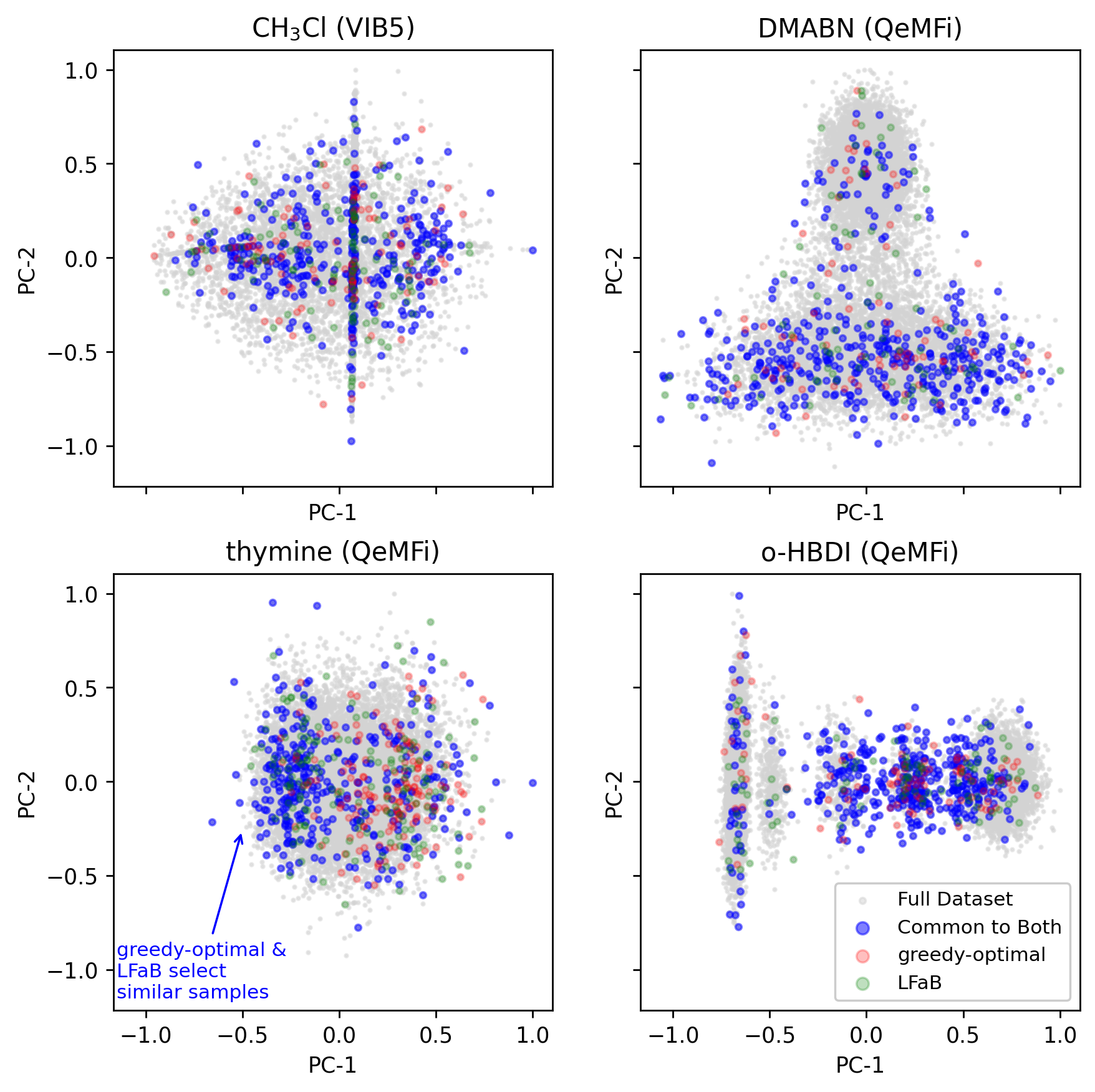}
    \caption{PCA scatter plot for selected molecules studied in this work indicating points selected by the greedy-optimal and LFaB methods in the first 500 iterations. Points selected by both methods are indicated separately. The axes are scaled to lie with unitary values for the two first principle component (PC-1 and PC-2) of the CM molecular descriptor. The LFaB method selects almost all the same data points as the greedy-optimal method. This indicates the novel LFaB method is as good as the greedy-optimal selection of training data, however by a using a lower (and thereby cheaper) fidelity as reference data.
    }
    \label{fig_SF_combinedPCA}
\end{figure}

Fig.~\ref{fig_SF_combinedPCA} shows the 2D PCA for $\rm CH_3Cl$ from VIB5, DMABN, thymine, and o-HBDI from QeMFi. The total AL training pool is shown in the foreground. For each molecule, three classes of samples are highlighted in order to compare greedy-optimal and LFaB methods. The first set is the collection of points that are selected during the first 500 AL iterations by both sampling methods. The second is those selected by the greedy-optimal method but not selected by LFaB, and third corresponds to the ones selected only by the LFaB method. For $\rm CH_3Cl$, the low fidelity is HF-ccpVTZ while for the molecules of QeMFi it is the use of the 321G basis set. 
In all 4 molecules it can be seen that LFaB selects nearly all the same molecular configurations as the greedy-optimal method. 
In contrast, Fig.~\ref{fig_combinedPCA} in the supplementary information compares the points selected by GPR variance and LFaB with significantly different points being selected by both these methods (see \ref{sec_PCA_SI}). This is a second indicator that the herein developed LFaB method performs as well as the ideal choice approach identifying the right areas of high uncertainty in the proxy molecular configuration space whereas the use of GPR variance picks vastly different samples.

\subsection*{Calibration Curves}
\begin{figure}[H]
    \centering
    \includegraphics[width=\linewidth]{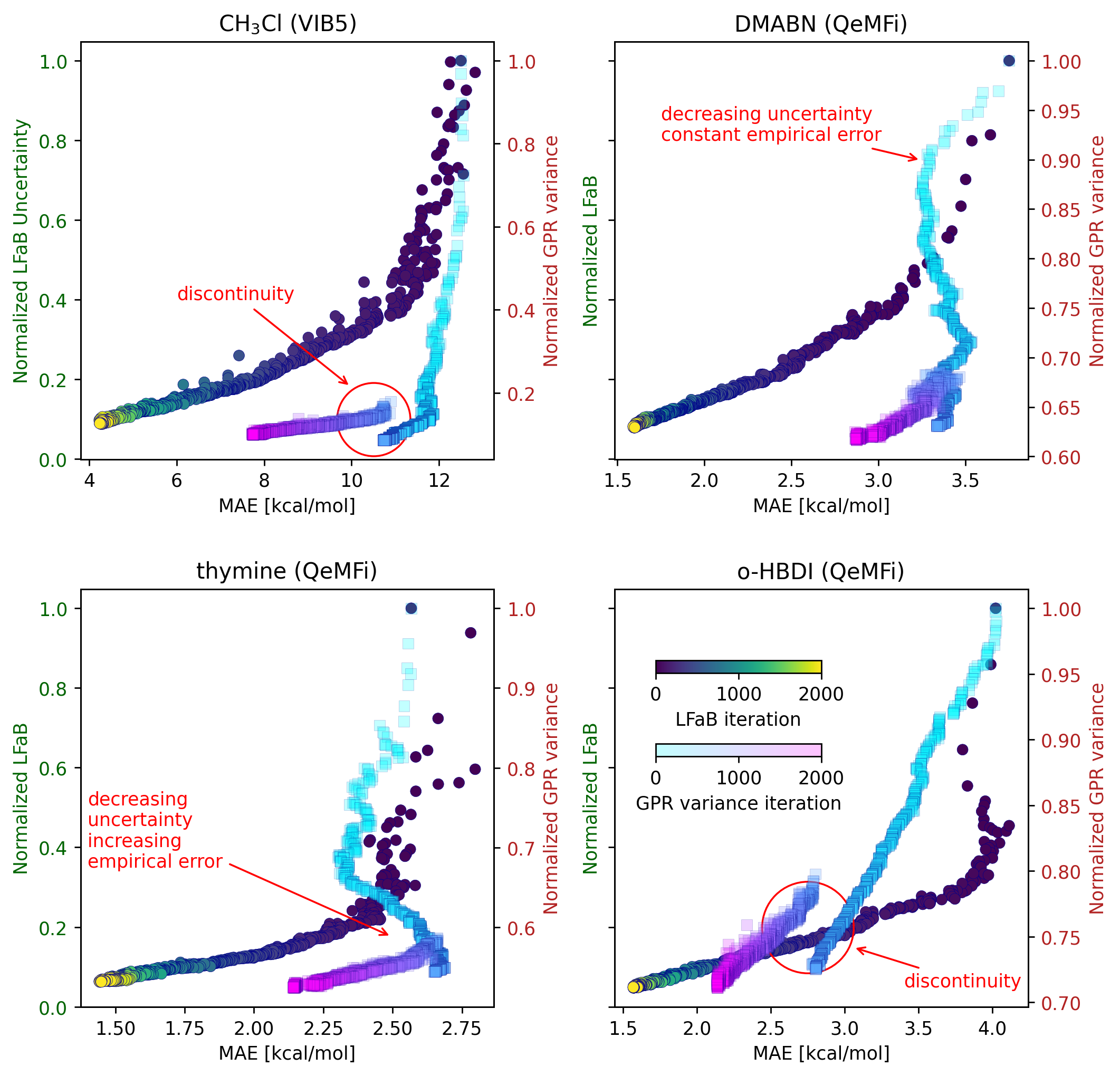}
    \caption{Uncertainty of the trained model versus empirical error, computed as MAE, over the holdout test set for selected molecules from the datasets used in this work. These curves are also referred to as \textit{calibration curves}. The color-scheme of the points correspond to the active learning iteration. Both the LFaB and GPR variance are reported in normalized units. The left axis corresponds to LFaB while the right axis corresponds to the GPR variance. The curves for $\rm CH_3Cl$ correspond to the prediction of \textit{ab initio} ground state PES while the QeMFi molecules correspond to the prediction of DFT excitation energies. It is observed that the LFaB method results in a systematic decrease in both uncertainty and MAE while the GPR variance shows stagnation of MAE and discontinuities in the uncertainty.}
    \label{fig_calibration_combined}
\end{figure}

For the same four molecules as before, Fig.~\ref{fig_calibration_combined} reports the calibration curves, a plot of empirical error against the uncertainty of the trained model. In this case, the uncertainty of the model is either the variance of the model or the bias (in case of LFaB). 
The horizontal axis reports the MAE in kcal/mol, the left vertical axis notes the normalized value of bias from LFaB, and the right vertical axis reports the normalized GPR variance. The normalization is carried out by dividing the set of bias or variance values by their respective maximum. 

It can be observed that for all four molecules, the empirical error decreases with a decrease in the LFaB for each step of the AL iteration (indicated by the color of the markers). This decrease is steady and without any discontinuities meaning that LFaB was able to identify the correct molecular configurations with highest bias in order to translate that into a lower empirical error in the trained GPR model. One the other hand the GPR variance results in cases with discontinuities ($\rm CH_3Cl$ and o-HBDI), constant empirical errors for decreasing uncertainty (DMABN), and in the case of thymine an increase in empirical error for a decrease in uncertainty. These results once again confirm the superiority of the proposed LFaB method over conventionally used variance based sampling methods. The model ensemble variance is not depicted in these plots since the empirical error of this approach is in all cases as large as that for the model built with the GPR variance for sampling. 

\section*{Conclusion}\label{sec_conclusion}
Existing sampling techniques in active learning employ variance as the primary driver in selection of training data. These are often seen to be limiting in their effectiveness. In this work, we introduced a novel sampling scheme that employed the bias of the model computed with a low (and thereby cheaper to acquire) fidelity. The low-fidelity-as-bias scheme was shown to be superior to variance based sampling methods such as the use of GPR model variance or model ensemble variance. In fact, the LFaB method was shown to perform as well as the best case greedy-optimal sampling method but without the high cost associated with computing high fidelity labels \textit{a priori}. The LFaB method was benchmarked on the prediction of several QC properties including excitation energies and shown to be a better strategy of selecting training samples than random sampling. Training data set size reduction of a factor of 5 was observed with the use of LFaB. In some cases this reduction was larger.

One major challenge in the iterative AL scheme used in this work is the repeated predictions over the training pool. For a very large dataset with a few million data-points, this would often become regressive. Of course, this is a challenge with any AL scheme and not limited to the LFaB method. A possible option to overcome such restrictions is the use of low rank approximations for kernel based methods \cite{williams2006gaussian}.
An extension of this work could involve applying the LFaB method to NN-architecture based ML models for QC such as SchNET \cite{schutt2018schnet, schutt21a_PAINN} or ANI \cite{ANI-NN_ML, smit17_ANI-1}. 

Overall, this work has implemented an effective sampling strategy in the active learning selection of training data for the molecular configuration space. The LFaB method developed in this work is a robust tool in the use of ML for QC which is easy to implement and shows effective reduction of both training set size and empirical error.

\newpage
\section*{Figures}\label{figure_section}

\newpage
\section*{Methods}\label{sec_methods}
This section introduces methods employed in this work including the different variance based sampling methods that are studied. Details about datasets and molecular descriptors used are presented in the Supplementary Information section \ref{app_methods}.

\subsection*{Gaussian Process Regression}\label{sec_GPR_methods}
Consider the training dataset denoted by $\mathcal{T}:=\{(\boldsymbol{x}_i,y_i)\}_{i=1}^{N_{\rm train}}$ where the inputs are $\boldsymbol{x}_i\in \mathcal{X}\subseteq\mathbb{R}^D$ and outputs are $y_i=g(\boldsymbol{x}_i)\in\mathbb{R}$ computed at some fidelity $f$. 
One can collect all the input features from the training dataset into $\boldsymbol{X}$ and the corresponding outputs into $\boldsymbol{Y}$.
The Gaussian covariance function is defined as 
\begin{equation}
    k(\boldsymbol{x},\boldsymbol{x}') = \exp{\left(-\frac{\lVert \boldsymbol{x}-\boldsymbol{x}'\rVert_2^2}{2\rho^2}\right)}~,
\end{equation}
where the term $\rho$ is a length-scale parameter for the covariance function. The covariance function is often also called the \textit{kernel} function. 
Gaussian Process Regression (GPR) is a \textit{non-parametric} ML method which assumes that the relationship between the input to the ML model and the predicted output is a probability distribution over functions defined by a Gaussian process (GP) \cite{williams2006gaussian}. 
The relationship between $\boldsymbol{X}$ and $\boldsymbol{Y}$ can be modeled as 
\begin{equation}
    \boldsymbol{Y}=g_{ML}(\boldsymbol{X}) + \boldsymbol{\epsilon}~,
\end{equation}
with $\boldsymbol{\epsilon}\sim\mathcal{N}(0,\sigma^2_{N_{\rm train}})$, that is, the error is Gaussian distributed with zero mean and constant variance dependent on the training data.

The prediction from a GPR model for a query input feature $\boldsymbol{x}_q$ is given as 
\begin{equation}
    g_{ML}(\boldsymbol{x}_q) = \boldsymbol{K}_q (\boldsymbol{K}+\sigma^2_{N_{\rm train}}\boldsymbol{I})^{-1}\boldsymbol{Y}~,
    \label{eq_GPR_mean}
\end{equation}
where $(\boldsymbol{K}_q)_i = k(\boldsymbol{x}_q,\boldsymbol{x}_i)$, and $(\boldsymbol{K})_{ij}=k(\boldsymbol{x}_i,\boldsymbol{x}_j)$ with $i$ and $j$ indexing the input features in $\mathcal{T}$. Further, $\sigma$ is a regularization parameter. 
In this work, the GPR process is implemented using the \texttt{GPyTorch} library for python \cite{gpytorch}.

\subsection*{Existing Sampling Techniques for Active Learning}\label{sec_uncertainty_measure}
AL schemes attempt to select the optimal training samples for a ML model based on the model's uncertainty or error measure, hereon denoted as $\phi$. 
In such a setup, there exists a pool of all possible candidates, denoted as $\boldsymbol{x}\in\mathcal{U}$, from which an optimal training dataset, $\mathcal{T}$, must be computed. First, a small number of candidates are randomly selected from $\mathcal{U}$ for which labels are computed as $y_i = g(\boldsymbol{x}_i)$ which results in the initial $\mathcal{T}:=\{(\boldsymbol{x}_i,y_i)\}_{i=1}^N$. This is used to train a ML model and used to compute the uncertainty of the model over the remaining candidates in $\mathcal{U}$. 
The optimal selection for the current iteration of AL is carried out with 
\begin{equation}
    \boldsymbol{x}_{\rm optimal} = \argmax_{\boldsymbol{x}\in\mathcal{U}}\phi(\boldsymbol{x})~.
\end{equation}
Once selected, the label for this sample is computed as $y_{\rm optimal}=g(\boldsymbol{x}_{\rm optimal})$ and this pair is added to the training dataset: $\mathcal{T}\cup\{(\boldsymbol{x}_{\rm optimal},y_{\rm optimal}))\}$. This process is iteratively repeated until a certain training set size is reached. Algorithm \ref{alg:uncertainty_sampling} in section \ref{sec_AL_details} delineates the above procedure for AL. Below, common sampling strategies are detailed.

\paragraph*{Random Sampling}
In general ML approaches, the most common method to select training samples is by randomly sampling across $\mathcal{U}$ to build the training dataset $\mathcal{T}$. 
Although this approach is not a bias or variance informed method, it is still relevant to use this process as a comparison in order to evaluate the results of this work. 

\paragraph*{GPR Variance}
Consider the query input, $\boldsymbol{x}_q$. 
For a GPR model described in \nameref{sec_GPR_methods}, 
the standard deviation of the predictive distribution for $\boldsymbol{x}_q$ is given by 
\begin{equation}
\phi_{\rm \text{GPR}}\left(\boldsymbol{x}_q\right) = k(\boldsymbol{x}_q,\boldsymbol{x}_q) - \boldsymbol{K}^{\top}_q \bigl(\boldsymbol{K} + \sigma_n^2\boldsymbol{I}\bigr)^{-1}\boldsymbol{K}_q
\label{eq_gpr_variance}
\end{equation}
where $(\boldsymbol{K}_q)_i=k\left(\boldsymbol{x}_q, \boldsymbol{x}_i\right)$ with $i$ indexing the input features in $\mathcal{T}$. Based on the above formulation, it can be inferred that the value of $\phi_{\rm \text{GPR}}$ will be higher when $\boldsymbol{x}_q$ is `farther' from the $\boldsymbol{x}_i$ in the training dataset.

\paragraph*{Model Ensemble Variance}
The bootstrap aggregation ensemble approach is another common way to selection samples in AL, particularly for neural network based architectures \cite{bradley_bootstrap_1992, bishop_book_ML_2006}. The notion is to first train an ensemble of $n$ ML models trained on randomly selected subsets of the training data. The variance for a query input $\boldsymbol{x}_q$ is computed as 
\begin{equation}
    \phi_{n}(\boldsymbol{x}_q) = \frac{1}{n-1}\sum_{l=1}^{n}\left(g_{ML}^{(l)}(\boldsymbol{x}_q)-g_{\rm avg}(\boldsymbol{x}_q)\right)^2~,  
    \label{eq_bootstrap_UQ}
\end{equation}
where 
\begin{equation}
    g_{\rm avg}(\boldsymbol{x}_q) = \frac{1}{n}\sum_{l=1}^{n}g_{ML}^{(l)}(\boldsymbol{x}_q)~,
\end{equation}
that is, the mean prediction of the different models, $g_{ML}^{l}$ in the ensemble. Often the standard deviation version of Eq.~\eqref{eq_bootstrap_UQ} is used. 
Note that the division by $N_{\rm ensemble}-1$ ensures this is an unbiased estimator for the standard deviation, particularly for small $N_{\rm ensembles}$. 
In this work, a 5-fold ensemble is used, that is 5 different ML models are trained. For the ensemble variance method, the built ML models need to have some difference in the training data that is used. This is often achieved by making the training subset size smaller than the size of $\mathcal{T}$. This introduces another nuance while comparing this approach to others. The number of training samples used for each of the models in the ensemble is less than the ones used for the other sampling methods. In order to minimize the effect of this reduction in number of training samples while retaining sufficient difference in the ensemble of models that are trained, in this benchmark, each model of the ensemble uses 85\% of the total training set available during each AL iteration. This ensures that the ML models trained in this scheme do not suffer significantly in terms of number of training samples in comparison to the other sampling methods which have access to 100\% of the training samples to build the GPR model.

\paragraph*{Greedy-Optimal Sampling}
The greedy-optimal sampling explicitly requires that the labels $y=g(\boldsymbol{x})$ be computed for all $\boldsymbol{x}\in\mathcal{U}$. This sampling is studied here as the best case sampling scheme. The argument is that if one had complete knowledge of how the function $g$ looked like, how would one go about selecting training samples. 
Formally, consider that the reference output is computed for each $\boldsymbol{x}\in\mathcal{U}$ as $g(\boldsymbol{x})$. Then for some query input feature $\boldsymbol{x}_q$ is defined as 
\begin{equation}
    \phi_{\rm greedy}(\boldsymbol{x}_q)=\lvert g_{ML}(\boldsymbol{x}_q)-g(\boldsymbol{x}_q) \rvert~.
\end{equation}

\subsection*{Low-fidelity as Bias}\label{lfab_method}
The overall error of a ML model is often written down as 
$$
    \rm Error=Bias^2+Variance+\text{Irreducible Error}~,
$$
that is, it depends on both the variance and bias of the model \cite{Geman1992NeuralNA,Domingos2000AUB}. Previously mentioned sampling techniques attempt to reduce the variance of the ML models by selecting samples from the training pool with the highest variance. It was identified by some of us earlier that the ML models built for quantum chemical properties do not reduce the empirical error even with a reduction in the variance of the model \cite{holzenkamp2025_uncertainty_GPR}. In this current work, we assume that the dominating component of the error is the bias of the ML models.

In order to ensure a reduction in model error, it is therefore pertinent that training data sampling schemes address the reduction of the bias. The challenge here is that computing the bias of the model requires the computation of $y=g(\boldsymbol{x})$ for all candidates in $\mathcal{U}$, the actual labels. This is precisely the principle behind the greedy-optimal sampling discussed above.
However, this renders the use of ML in QC redundant since the entire pool of data has the property of interest already computed.

To reduce the bias of the ML model without incurring redundancy, we propose the use of labels $y_{\rm low}=g^{\rm low}(\boldsymbol{x})$ computed at a lower fidelity. These labels are much cheaper to acquire due to $g^{\rm low}$ being a a cheaper to perform computation than $g$. This low fidelity label is used to approximate the bias of the ML model trained at the target fidelity. This method is referred to as Low-Fidelity-as-Bias (LFaB). 

In several applications, particularly QC, lower fidelities are orders of magnitude cheaper to compute than the higher fidelities \cite{Crawford_CCSD_theory_2000, szabo2012modern_book}. This makes them an attractive proxy for the approximation of model bias.
For a query descriptor, $\boldsymbol{x}_q$, the LFaB is computed as the absolute difference in prediction and reference of the label at the lower fidelity:
\begin{equation}
    \phi_{\rm LFaB}(\boldsymbol{x}_q) = \left\lvert g_{ML}^{\rm low}(\boldsymbol{x}_q)-g^{\rm low}(\boldsymbol{x}_q)\right\rvert~.
    \label{eq_LFaB}
\end{equation} 
By construction, this requires the labels $g^{\rm low}(\boldsymbol{x})$ for $\boldsymbol{x}\in\mathcal{U}$ to be computed only for the lower fidelity making it a much cheaper measure of the bias in comparison to greedy-sampling. 
The procedure to use LFaB to adaptively select training samples is as follows:
\begin{enumerate}
    \item Build initial training dataset $\mathcal{T}$. 
    \item For the molecular configurations in $\mathcal{T}$ also compute the low fidelity labels to make the low fidelity training dataset $\mathcal{T}^{\rm low}:=\{(\boldsymbol{x}_i,y_{i,\rm low})\}$.
    \item Compute the low fidelity labels $y_{\rm low}=g^{\rm low}(\boldsymbol{x})_{i=1}^{N}$ for $\boldsymbol{x}\in\mathcal{U}$.
    \item Train two separate GPR models $g_{ML}$ and $g_{ML}^{\rm low}$ on the two respective datasets.
    \item Make predictions over $\mathcal{U}$ with $g_{ML}^{\rm low}$.
    \item Compute LFaB using Eq.~\eqref{eq_LFaB}.
    \item Pick $\boldsymbol{x}_{\rm optimal}$ as $\boldsymbol{x}\in\mathcal{U}$ that has the highest value of $\phi_{\rm LFaB} (\boldsymbol{x})$ and compute both $y_{\rm optimal}=g(\boldsymbol{x}_{\rm optimal})$ and $y_{\rm optimal,low}=g^{\rm low}(\boldsymbol{x}_{\rm optimal})$.
    \item Increment the training data as $\mathcal{T}^F\cup\{(\boldsymbol{x}_{\rm optimal},y_{\rm optimal})\}$ and $\mathcal{T}^{\rm low}\cup\{(\boldsymbol{x}_{\rm optimal},y_{\rm optimal,low})\}$ respectively.
    \item Retrain the models $g_{ML}$ and $g_{ML}^{\rm low}$ and repeat from step 4 iteratively.  
    
\end{enumerate}

\subsection*{Error Metrics}
ML models that are produced as a result of the different AL schemes proposed in this work are evaluated on the basis of mean absolute error (MAE) over a holdout test set $\mathcal{V}:=\{(\boldsymbol{x}_t,y^F_t)\}_{t=1}^{N_{\rm test}}$ where $y^F_t=g^F(\boldsymbol{x}_t)$ is computed at the target fidelity $F$. This test set is not used in any of the training phases. For each case, the MAE is computed using a discrete $L_1$ norm as follows:
\begin{equation}
    \mathrm{MAE} = \frac{1}{N_{\rm test}}\sum_{t=1}^{N_{\rm test}}\left\lvert g_{ML}\left(\boldsymbol{x}_t^{ \rm test}\right) - y_t\right\rvert~.
    \label{eq_MAE}
\end{equation} 
The MAE is always computed with reference to the target fidelity QC properties. The evolution of MAE with increasing model complexity is studied as the learning curve and indicates how well an ML model is able to predict over unseen data \cite{cortes1993learning}. For GPR, model complexity is explicitly dependent on number of training samples used. Therefore, in this work, MAE versus number of training samples used are studied in order to make inferences about the ability of a GPR model in predicting properties for inputs from $\mathcal{V}$.

\backmatter

\section*{Code and Data availability}
The programming scripts used for this work can be openly accessed at the \href{https://github.com/SM4DA/LFaB}{LFaB} GitHub repository. All data used in this work comes from openly available datasets which are appropriately cited.

\section*{Supplementary information}
Supplementary sections S1-S2, figures SF1-SF4, algorithm SA1.

\section*{Acknowledgments}
The authors acknowledge support by the DFG through the project ZA 1175/3-1 as well as through the DFG Priority Program SPP 2363 on “Utilization and Development of Machine Learning for Molecular Applications – Molecular Machine Learning” through the project ZA 1175/4-1. The authors would also like to acknowledge the support of the `Interdisciplinary Center for Machine Learning and Data Analytics (IZMD)' at the University of Wuppertal.

\section*{Declarations}
The authors declare that there is no conflict of interest. 

\bibliography{main}

\pagebreak
\begin{center}
\textbf{\Large Supplementary Information for LFaB}
\end{center}
\vspace{1cm}
\setcounter{equation}{0}
\setcounter{section}{0}
\setcounter{figure}{0}
\setcounter{table}{0}
\setcounter{page}{1}
\makeatletter
\renewcommand{\theequation}{S\arabic{equation}}
\renewcommand{\thefigure}{SF\arabic{figure}}
\renewcommand{\thepage}{S\arabic{page}} 
\renewcommand{\thesection}{S\arabic{section}}  
\renewcommand{\thetable}{S\arabic{table}} 
\renewcommand{\thealgorithm}{SA\arabic{algorithm}} 

\section{Supplementary Methods}\label{app_methods}

\subsection{Active Learning}\label{sec_AL_details}
\begin{algorithm}[htb!]
\caption{Training data sampling with Active Learning.}
\begin{algorithmic}[1]
\Require AL pool of input features $\mathcal{U}$, number of AL iterations $n_{\rm iter}$, initial number of training samples $n_{\rm init}$
\State $\mathcal{A}\leftarrow$Randomly select $n_{\rm init}$ samples from $\mathcal{U}$
\State $\boldsymbol{Y}\leftarrow y_i=g(\boldsymbol{x}_i),~\forall \boldsymbol{x}_i\in\mathcal{A}$ \Comment{Compute labels for initial data}
\State $\mathcal{U}\leftarrow\mathcal{U}\setminus\mathcal{A}$
\State $\mathcal{T}\leftarrow\{(\boldsymbol{x}_i,y_i)\}_{i=1}^{n_{\rm init}}$ for $\boldsymbol{x}_i\in\mathcal{A}$ and corresponding $y_i\in\boldsymbol{Y}$ 
\State Train initial model $g_{ML}$ on $\mathcal{T}$
\For{$i = 1$ to $n_{\rm iter}$}
    \State $\Phi\leftarrow[\phi(\boldsymbol{x_q})]_{\boldsymbol{x}_q\in\mathcal{U}}$ \Comment{Compute bias or variance of model $\phi$}
    \State $\rm \boldsymbol{x}_{\rm optimal}\leftarrow\arg\max_{\boldsymbol{x}\in\mathcal{U}}\Phi(\boldsymbol{x})$
    \State $y_{\rm optimal}\leftarrow g(\boldsymbol{x}_{\rm optimal})$
    \State $\mathcal{A}\leftarrow\mathcal{A} \cup \boldsymbol{x}_a$
    \State $\mathcal{U}\leftarrow\mathcal{U}\setminus\mathcal{A}$
    \State $\mathcal{T}\cup\{(\boldsymbol{x}_{\rm optimal},y_{\rm optimal})\}$ \Comment{Update training set}
    \State Retrain $g_{ML}$
\EndFor
\State \textbf{return} Trained model $g_{ML}$
\end{algorithmic}
\label{alg:uncertainty_sampling}
\end{algorithm}
Active learning (AL) is a class of methods in machine learning (ML) that attempt to make a well informed decision on how to select optimal training samples from a training pool of unlabeled data.
In the setup used in this work, one has several molecular configurations for which computing the quantum chemistry (QC) labels is expensive. One wishes to reduce the total number of such calculations by letting AL schemes identify the best training samples from the pool. The overarching algorithm is shown in Algorithm \ref{alg:uncertainty_sampling}. The different bias or variance based sampling methods are discussed in the main text.

\subsection{Datasets}\label{sec_datasets}
This work employs several datasets in order to assess the effect of using LFaB over sampling methods for active learning. For completeness, key details of these datasets are mentioned here.

\subsubsection{QM7b}\label{sec_qm7_data}
The QM7b dataset is a collection of 7211 molecules with several heavy atoms including C, N, and O \cite{montavon2013machine}. The dataset provides the atomization energies of these molecules, that is, the measure of energy required to break a molecule into unbound atoms. The energies are calculated with the Hartree Fock (HF) \cite{1928_SCF_HF_slater,Blinder_2019_HF_chapter}, Møller–Plesset perturbation theory (MP2) \cite{Quin_MP2_DFT_theory_2005, Yost_MP2_theory_2018, Pogrebetsky_MP2_theory_2023} \cite{Purvis_CCSD_theory_1982, Bartlett_CCSD_theory_2007, Crawford_CCSD_theory_2000}, and Coupled Cluster Singles and Doubles perturbative Triples (CCSD(T)). For each method, three basis set sizes are used for computation, STO3G, 631G, and ccpVDZ. Therefore, there are a total of 9 point calculations made for each molecule in this dataset. The most expensive method, or fidelity, is the CCSD(T) method with the cc-pVDZ basis set, hereon denoted as CCSD-ccpVDZ. Similarly, each method and basis set is hereon concatenated in reference such as HF-STO3G. 

\subsubsection{VIB5}\label{sec_vib5_data}
The VIB5 database contains high accuracy \textit{ab initio} potential energy surfaces (PES) for several small molecules such as $\rm CH_3Cl$ and $\rm CH_3F$ \cite{zhang_vib5_2022}. For these two molecules, there are 44,819 and 82,653 point geometries respectively. This was achieved by performing an energy-weighted Monte Carlo sampling strategy to internal coordinates of the molecules such as the X-C-H bond angles (X represents the halogen), dihedral angles, and bond lengths between C and constituent atoms.
Corresponding energies are computed with varying degrees of accuracy. The fidelities used in this work are the CCSD(T)-ccpVQZ, MP2-ccpVTZ, HF-ccpVQZ, and HF-ccpVTZ. The target fidelity assumed is the CCSD(T)-ccpVQZ for both molecules.

\subsubsection{QeMFi}\label{sec_qemfi_data}
The quantum chemistry multifidelity (QeMFi) dataset \cite{vinod_2024_QeMFi_zenodo_datatset, vinod2024QeMFi_paper} consists of nine chemically diverse molecules with varying sizes with QC properties computed with the DFT formalism \cite{vinod2024QeMFi_paper, vinod_2024_QeMFi_zenodo_datatset}. These molecules include 2-(methyliminomethyl)phenol (SMA), 4-(dimethylamino) benzonitrile (DMABN), and 4-(2-hydroxybenzylidene)-1,2-dimethyl-1H-imidazol-5(4H)-one (o-HBDI). For each of the nine molecules, excitation energies are computed with the DFT \textit{ansatz} using the CAM-B3LYP functional. The fidelities available in this dataset are ordered by the size of the basis set used, namely, STO3G, 321G, 631G, def2-SVP, and def2-TZVP. The energies are computed for $9\times 15,000$ geometries. In this work, the various AL strategies are assessed for individual molecules of the QeMFi dataset for different $f_b$ baseline fidelities. However, STO3G is omitted since it was identified in previous works with QeMFi as a fidelity with poor characteristics \cite{vinod2024_nonnestedMFML, vinod2024_gamma_curve_error_contours}. 

\subsection{Molecular Descriptors}\label{sec_representations}
Intermediate steps in the use of machine learing in quantum chemistry applications involve making the choice of \textit{molecular descriptors} or \textit{representations} which are mappings between the Cartesian coordinates of the molecular configurations to machine learnable input features \cite{bartok2013representingSOAP, schutt2014represent}. Often, QC details of the molecule of interest are also included such as the nuclear charge or tetrahedral bond angles. Several representations are actively used in the ML-QC workflow. These include Coulomb Matrices (CM) \cite{RupCM}, smooth overlap of atomic positions (SOAP) \cite{bartok2013representingSOAP}, the spectrum of London and Axilrod–Teller–Muto (SLATM) representation \cite{Huang2020slatm}, atom-centered symmetry functions (ACSFs) of Behler \cite{behler_2011_atomcentered, smit17_ANI-1,gastegger_2018_wacsf}, and the Faber-Christensen-Huang-Lilienfeld (FCHL) descriptor \cite{Christensen2020}.

One of the molecular descriptors used in this work is the simple unsorted Coulomb Matrix (CM) \cite{RupCM}. For a molecule whose atoms are indexed by $i$ with corresponding nuclear charge $Z_i$, these are computed as
\begin{equation}
C_{i,j}:=
    \begin{cases}
        \frac{Z_i^{2.4}}{2}~,&i=j\\
        \frac{Z_i\cdot Z_j}{\left\lVert \boldsymbol{R}_i-\boldsymbol{R}_j\right\rVert}~,&i\neq j~,
    \end{cases}
    \label{CM_eq}
\end{equation}
where the Cartesian coordinates of an atom are given by $\boldsymbol{R}_i$ and the computation is performed for all pairs of atoms. The CM is a symmetric representation and therefore is of size $n(n-1)/2$ for a molecule with $n$ atoms. The experiments for QeMFi and VIB5 are performed using this descriptor.

For the atomization energies of the QM7b dataset, since there are several different molecules as opposed to a single molecule as for the case of VIB5 or QeMFi, the unsorted CM descriptors fail in terms of permutational invariance. Furthermore, the use of CM matrices sorted by their row norm is known to introduce discontinuities in the learned function \cite{RupCM, krae20a}. Therefore, this descriptor is avoided. Instead, based on previous work with this dataset in refs.~\cite{zasp19a, vinod_2024_oMFML} the SLATM representation \cite{Huang2020slatm} is used. 

\section{Supplementary Results}\label{app_sup_results}
Additional results that accompany the main text are presented in this section. 
\begin{figure}[t]
    \centering
    \includegraphics[width=0.75\linewidth]{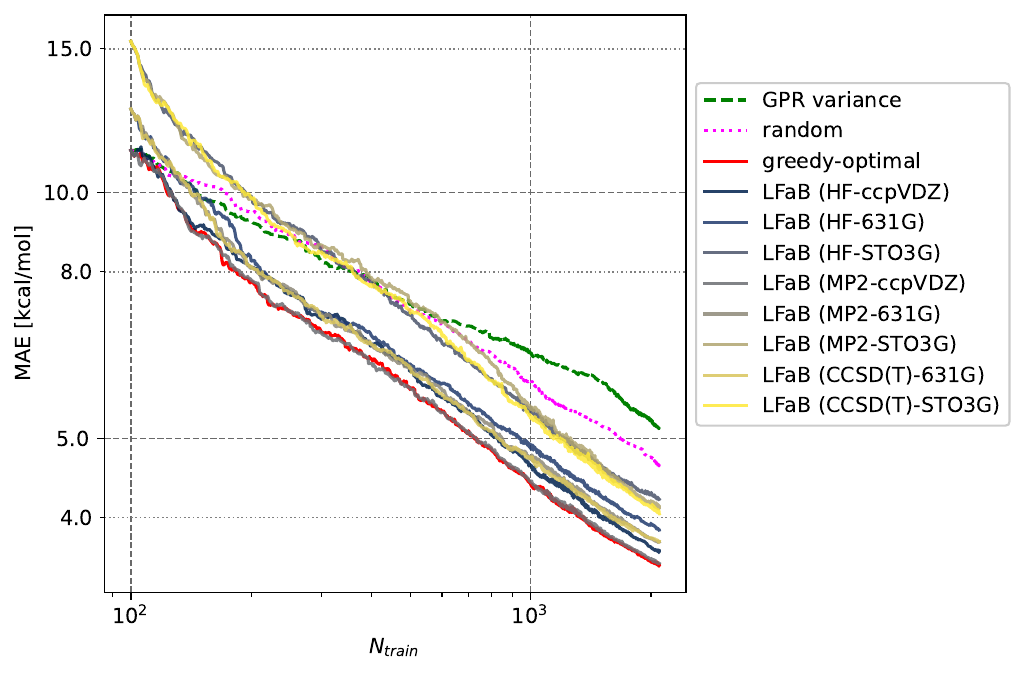}
    \caption{Learning curves for the prediction of atomization energies of the QM7b dataset where the target fidelity is fixed as CCSD(T)-ccpVDZ and the remaining fidelities are used in the LFaB method.}
    \label{fig_QM7b_fullrange}
\end{figure}
\subsection{Composite fidelity Hierarchy}\label{sec_composite_QM7b}
Another comprehensive way to test the robustness of the LFaB approach is shown in Fig.~\ref{fig_QM7b_fullrange}. In this case, the most expensive fidelity is CCSD(T)-ccpVDZ. Every other fidelity such as MP2-631G or HF-STO3G are considered cheap fidelities. The aim with LoFiAL here is to see how much the choice of QC method \textit{and} basis set have on the MAE of the model. Once again, the comparison with GPR-variance, random sampling, and greedy-optimal method is shown. Note that this time around, the notion of a fidelity is a combination of both the method and the basis set choice. While the LFaB method with cheaper fidelities does indeed provide a better MAE than the GPR-variance and random sampling approach, there is a varying degree of improvement that is seen as the chosen lower fidelity gets farther away from CCSD(T)-ccpVDZ. An observation can be made that the basis set choice seems to make more of a difference than the QC method chosen. For instance, the model built with LFaB (MP2-ccpVDZ) performs much better than that built with LFaB (CCSD(T)-STO3G). This observation can be understood better from the perspective that the choice of a basis set is the notion of where the wave function is truncated. The choice of truncation could potentially play a stronger role in the LFaB method than the specific QC method. Regardless, it is observed that the LFaB method is able to get close to the greedy-optimal sampling method in this case.

\subsection{PCA for GPR-variance} \label{sec_PCA_SI}
\begin{figure}[htb!]
    \centering
    \includegraphics[width=\linewidth]{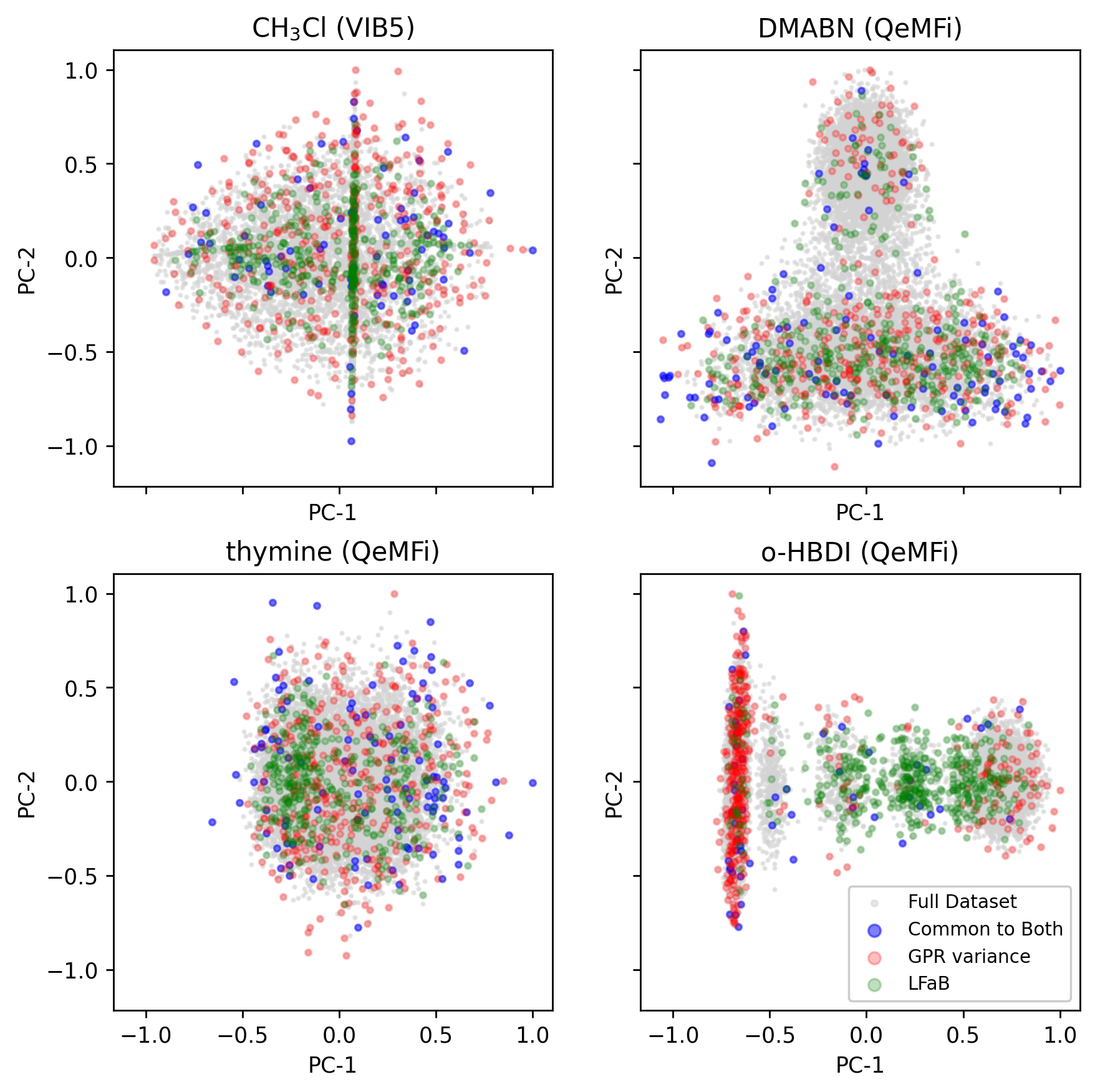}
    \caption{PCA scatter plot for selected molecules studied in this work indicating points selected by GPR-variance and LFaB in the first 500 iterations of active learning. Points selected by both methods are indicated separately. The axes are scaled to lie with unitary values for each principle component. The GPR-variance based AL places more importance to the peripheral regions of the scatter plots while the LFaB method shows more concentrated selection of data points across all the four molecules.}
    \label{fig_combinedPCA}
\end{figure}

Fig.~\ref{fig_combinedPCA} depicts the PCA scatter plot for 3 specific molecules: $\rm CH_3Cl$ from VIB5, and nitrophenol and DMABN from QeMFi. For these molecules, the 2-D PCA of the entire 15,000 samples is depicted in the background. The scatter plots include three distinct groups of points. Those data points selected using GPR-variance, those selected using the LFaB method, and finally those common to both. Only the first 500 samples that were chosen, that is the first 500 iterations of AL, are shown in order to have a less cluttered view of the PCA plots. The PCA plots are only shown for the AL pool and not for the initial training data and holdout test set. 

In all three cases, it can be observed that there are rather few samples that are common to the two sampling schemes. The two methods do indeed seem to sample distinct parts of this proxy space. Consider the case for $\rm CH_3Cl$. The LFaB (HF-TZ) method is shown. The GPR-variance sampling approach has an almost even sampling of the proxy space with points at the periphery of the scatter plot being strongly considered. In contrast, the LFaB approach seems to concentrate in the $+$ shaped central beams with some deviations around this structure. The peripheral structure of the proxy space does not seem to be of great importance to this selection strategy. 

For the case of nitrophenol, as seen in the middle scatter plot of Fig.~\ref{fig_combinedPCA}, the PCA depicts 2 major clusters with 3 minor clusters in between them. The GPR-variance based sampling chooses sparingly on the left-side cluster and evenly from the right-most cluster. Once again this method picks data points that are at the edge of the proxy-space. In contrast, the LFaB method, here shown for the 321G fidelity performs sampling concentrated at the left side cluster with a narrower sampling on the right-most cluster.  A similar observation is made for DMABN where the LFaB (321G) method samples less on the top cluster in comparison to the GPR-variance sampling. 

\subsection{Batch active learning with LFaB}\label{SI_batch_addition}
\begin{figure}[htb!]
    \centering
    \includegraphics[width=0.9\linewidth]{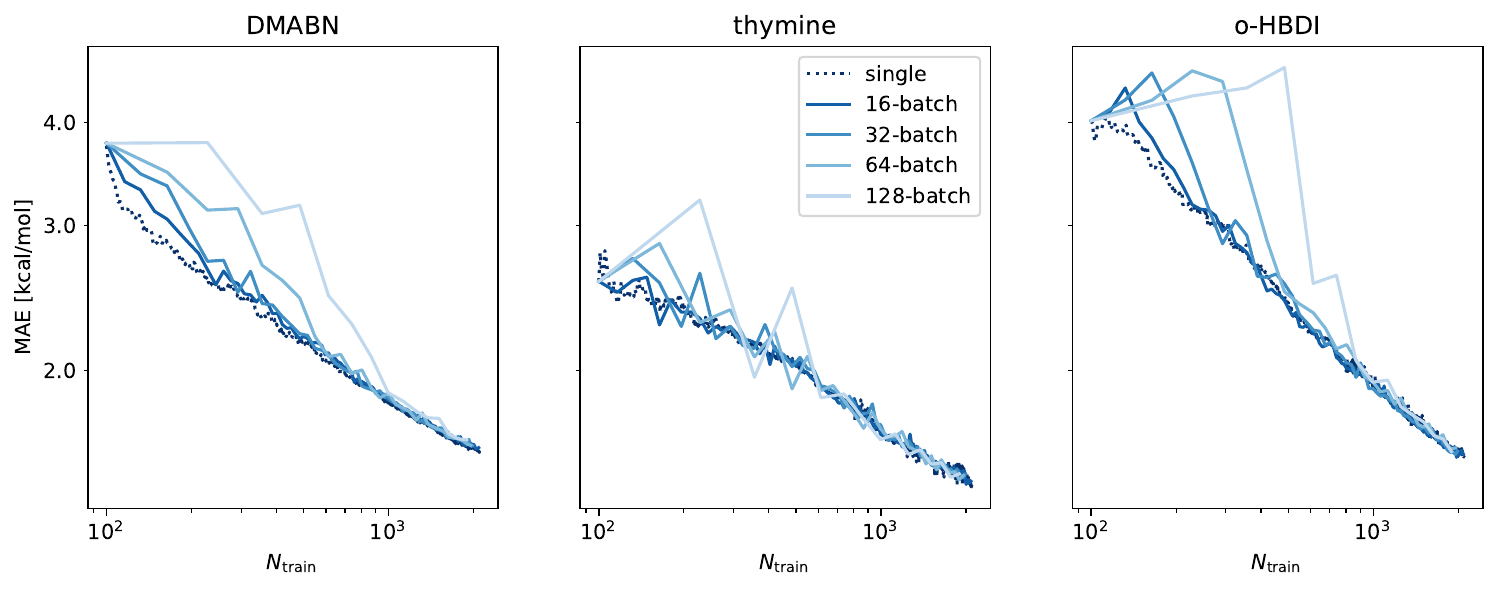}
    \caption{MAE vs training set size for the LFaB sampling method with batch-additions of training data compared with single data point addition. It can be seen that the addition of the training data in batches converges to the same MAE as the single point addition as shown in the main text.}
    \label{fig_batchsizetest}
\end{figure}
In the main text, the results for the different active learning sampling schemes are shown for the addition of one training data point during each iteration of active learning. While this is a good strategy to benchmark and study the methods, in practice, one is often interested to add training samples in batches.

In Fig.~\ref{fig_batchsizetest} the use of different batch sizes to add training data using LFaB is shown for three molecules of the QeMFi dataset for the prediction of excitation energies. It can be seen that the learning curves that report the MAE as a function of training samples converge regardless of the size of the batch.

\begin{figure}[htb!]
    \centering
    \includegraphics[width=0.9\linewidth]{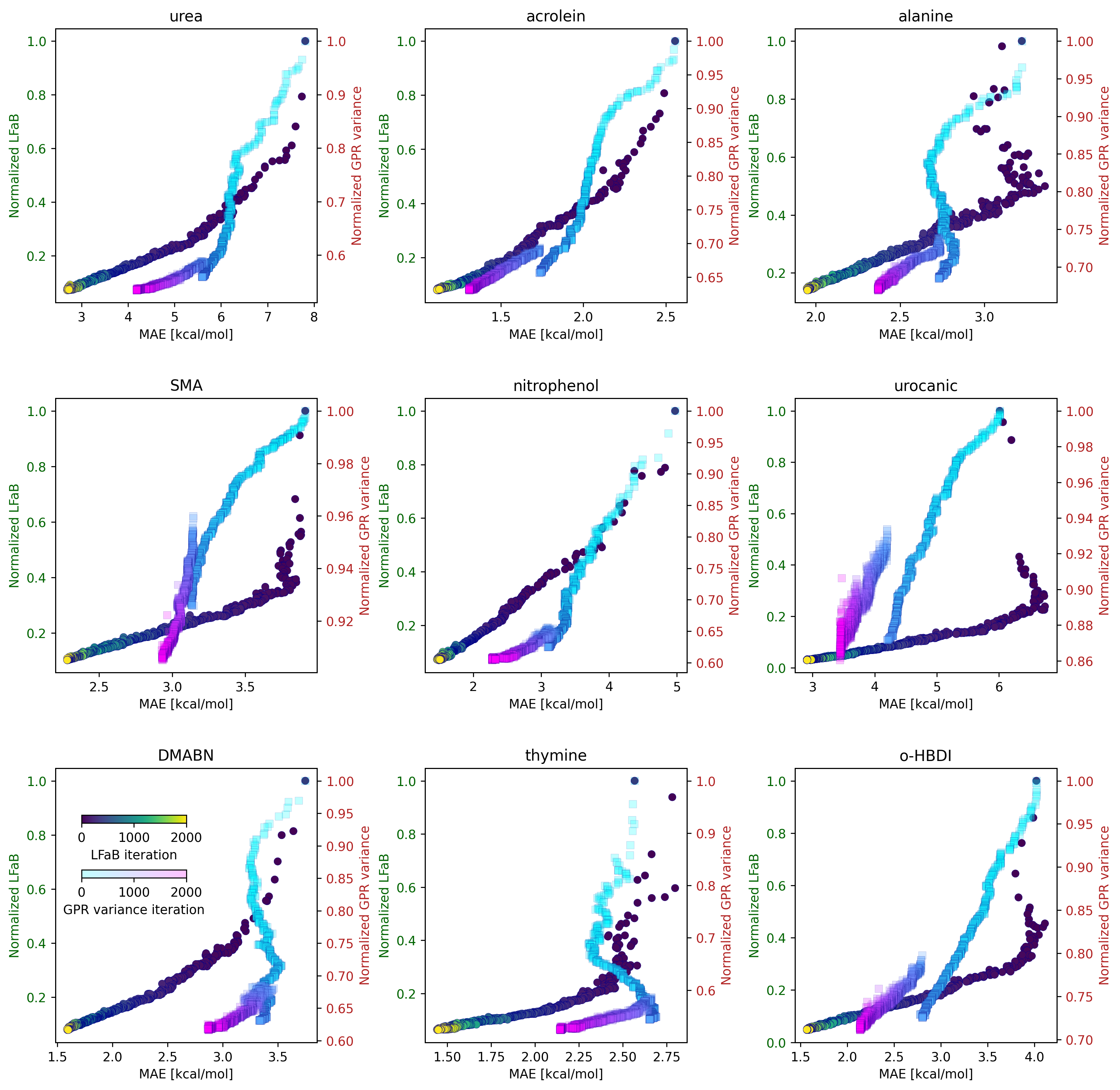}
    \caption{Calibration curves for prediction of excitation energies of the molecules from the QeMFi dataset. Curves for LFaB and GPR variance based sampling are shown.}
    \label{fig_qemfi_calib_all}
\end{figure}

\subsection{Calibration Curves for other molecules of QeMFi}

For completeness, Figure \ref{fig_qemfi_calib_all} shows the calibration curves for all 9 molecules from the QeMFi dataset.

\end{document}